\documentclass[superscriptaddress,aps,showpacs,twocolumn,twoside,pre,10pt]{revtex4-1}
\usepackage{amsmath} 
\usepackage{amssymb}
\usepackage{graphicx}
\usepackage{color}      
\usepackage{hyperref}   
\def\ket#1{|#1\rangle}
\def\bra#1{\langle#1|}
\def\scal#1#2{\langle#1|#2\rangle}

\def\abs#1{\left\lvert#1\right\rvert}

\def\={\!=\!}
\def\>{\!>\!}
\def\<{\!<\!}
\def\-{\!-\!}
\def\+{\!+\!}

\def\abs#1{\left|#1\right|}

\newcommand{\ave}[1]{\left< #1 \right>}

\newcommand{\trace}[1]{{\rm Tr}\left\{ #1 \right\}}
\newcommand{\ii}{\mathrm{i}}

\begin{document}

\title{Collective performance of a finite-time quantum Otto cycle}

\author{Michal Kloc}
\email[E-mail address: ]{michal.kloc@unibas.ch}
\affiliation{Institute of Particle and Nuclear Physics, Faculty of Mathematics and Physics, Charles University, V\,Hole{\v s}ovi{\v c}k{\'a}ch 2, Prague, 18000, Czech Republic}
\affiliation{Department of Physics, University of Basel, Klingelbergstrasse 82, CH-4056 Basel, Switzerland}

\author{Pavel Cejnar}
\affiliation{Institute of Particle and Nuclear Physics, Faculty of Mathematics and Physics, Charles University, V\,Hole{\v s}ovi{\v c}k{\'a}ch 2, Prague, 18000, Czech Republic}

\author{Gernot Schaller}
\affiliation{Institut f{\"u}r Theoretische Physik, Technische Universit{\"a}t Berlin, D-10623 Berlin, Germany}

\date{\today}

\begin{abstract}
We study the finite-time effects in a quantum Otto cycle where a collective spin system is used as the working fluid.
Starting from a simple one-qubit system we analyze the transition to the limit cycle in the case of a finite-time thermalization.
If the system consists of a large sample of independent qubits interacting coherently with the heat bath, the superradiant equilibration is observed.
We show that this phenomenon can boost the power of the engine. 
Mutual interaction of qubits in the working fluid is modeled by the Lipkin-Meshkov-Glick Hamiltonian.
We demonstrate that in this case the quantum phase transitions for the ground and excited states may have a strong negative effect on the performance of the machine.
Reversely, by analyzing the work output we can distinguish between the operational regimes with and without a phase transition.

\end{abstract}

\maketitle

\section{Introduction}
Bringing the concept of heat engines to the quantum regime raised new questions on optimal working schemes for such machines~\cite{Ali79,Kos13,Kos17,Kos18,Cak16}. 
A significant amount of effort has been invested into the attempt to overcome some classical limitations using quantum features of the working fluid (WF) or the heat baths~\cite{Scu03,Ros14,Jar16,Nie18}.
Experimental realizations of such microscopic engines are already feasible these days.
Recently, successful implementations have been reported in Refs.~\cite{Ros16,Mas19} using trapped ions and in Ref.~\cite{Kla19} where negatively charged nitrogen vacancies in diamond were employed.
Numerous theoretical proposals have also been made using superconducting qubits~\cite{Nis07,Cam15,Mar16} or optomechanical systems~\cite{Zha14,Gel15}.

An important direction of research leads towards finite-time thermodynamics employed in the cycle~\cite{Gev92,Fel04,Rez06,Sch08,Esp10,Esp10b,Fel12,Bol12,Wan13,Wu14,Ins16,Aba16,Wie19}. 
In this case the WF is not kept in contact with the heat bath for sufficiently long time to be considered as fully thermalized before the next stroke takes place. 
Starting from an arbitrary initial state, after several cycles the engine reaches a stable mode of operation corresponding to a limit cycle in any quantum thermodynamic diagram.
Description of the transition period as well as the limit cycle itself is then useful to understand the properties of such an engine.
A complementary question is how long it takes for the system to reach thermal equilibrium (within a given tolerance) with the heat bath.
Obviously, knowing how to decrease the time needed for the thermalization (which we simply call \textit{thermalization time} througout the paper) could help in gaining more power~\cite{Cam16,Vro18,Nie18b,Wat19}.

Similarly, effects of the finite-time duration of the stroke need to be quantified also for the unitary parts of the cycle in which an internal parameter of the WF is varied~\cite{Bea16,Cak19}.
Very often the optimal working protocol is achieved by quantum adiabatic driving where no population transfers between the energy levels occur. 
In order to fulfill the adiabatic condition the evolution must become significantly slow if the system is driven across the point where the energy levels get very close to each other.  
In particular, this is the case in the systems with a quantum phase transition (QPT) where the energy gap between the ground state and the first excited state closes at the critical point in the thermodynamic limit~\cite{Sac11,Car11}.
A~similar scenario can take place among the excited states if a so-called excited-state quantum phase transition (ESQPT) is formed~\cite{Cej06,Cap08,Str16}.
Even in strictly finite systems where only precursors of these phenomena appear, their presence may lead to significant population changes and thus may have a negative effect on the amount of work extracted.
On the other hand, it has been recently reported that the presence of a QPT may also have a possitive effect on efficiency of the heat engine~\cite{Ma17,Cha18} so some conclusive statement is needed.

In this paper we aim at investigating the finite-time effects both in the thermalization strokes and in the evolution of the WF with some non-thermal parameter.
The heat engine will be driven through the quantum Otto cycle which is briefly described in Sec.~\ref{Sec:TechBack}.
In the same Section we also introduce the concept of reference temperature which we will employ to monitor the evolution of the system during the cycle.
In Sec.~\ref{Sec:NonInt} we start with a toy model of a single qubit where we analytically reproduce the evolution of the system in the plane \lq mean energy vs. reference temperature\rq .
Further, we model the WF by a large-spin system  where a significant decrease in thermalization time is observed and explained in analogy to superradiance.
We demonstrate a power boost in this case compared to an incoherent ensemble of a large number of mutually uncorrelated qubits.
Our results in this Section complement the previously reported ones on performance enhancement due to collective effects in quantum transport~\cite{Kar11,Vog11,Sch16} or in models of  quantum batteries with global interactions~\cite{Bin15,Cam17,Le18,Fer18,And19}.
Recent publications~\cite{Jar16,Vro18,Nie18b,Wat19} also show the benefits of cooperative many-body effects in context of quantum heat engines, namely in Refs.~\cite{Nie18b,Wat19}  a direct link to Dicke superradiance~\cite{Dic54} is made.
In this paper we elaborate this analogy in more detail, namely we identify an operational region where power scales as $N^2$ and show that the classical superradiant equations written down by Gross and Haroche~\cite{Gro82} can be generalized to the finite-temperature regime.

Finally, in Sec.~\ref{Sec:LMG} we consider the Lipkin-Meshkov-Glick (LMG) model~\cite{Lip65} in the cycle.
This model is well-known to exhibit both a QPT and ESQPTs (see for example Refs.~\cite{Gil78,Rib07,Cej15}) so varying its control parameters may take the system through the critical point.
Similarly to the prior Section, the cyclic evolution is monitored in the plane \lq mean energy vs. reference temperature\rq\ where the traces of a QPT are identified.
In the end we discuss the effect of criticality on the performance of the engine and put our results in context with other works on a similar topic.

\section{Technical background}
\label{Sec:TechBack}
Throughout the paper we work with the units $\hbar=1$ and $k_B=1$.
\subsection{Quantum Otto cycle}
\begin{figure}[t]
	\centering
  \includegraphics[width=0.65\linewidth, angle=0]{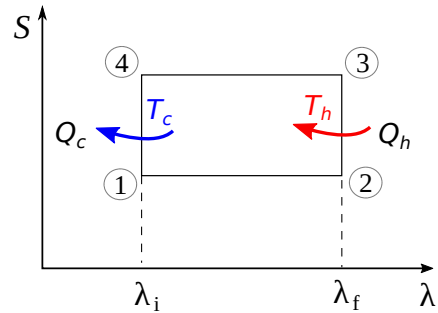}
	\caption{Schematic Otto cycle. $S$ is entropy  and $\lambda$ a non-thermal control parameter of the WF.}
	\label{fig1}
\end{figure}

The standard quantum Otto cycle consists of four strokes, see Fig.~\ref{fig1}.
The WF is initially in thermal equilibrium with the cold reservoir $T_c$, then it is decoupled from it and undergoes a unitary (thus isentropic) evolution with a non-thermal parameter $\lambda$ during the first stroke $1 \to 2$.
In the second stroke $2 \to 3$ it is brought into contact with the hot reservoir $T_h$ while the parameter $\lambda$ is fixed.
At the end of the stroke the WF is in thermal equilibrium with the hot bath.
In a similar way the WF reaches its initial state after the subsequent strokes $3 \to 4$ and $4 \to 1$.

Strictly speaking the perfect thermal equilibration is achieved in infinite time.
This idealized operational mode, however, harvests work at zero power.
So whenever we refer to any state as reaching thermal equilibrium (or being fully thermalized) in a finite time we implicitly mean \lq within a certain tolerance\rq .

Heat $Q_{h}$  injected into the WF during  $2\to 3$ and heat $Q_{c}$ released during  $4\to 1$ can be expressed as
 \begin{equation}
 Q_{h}=\trace{H_{\rm f} (\rho_3-\rho_2)}, \ \ \ Q_{c}=\trace{ H_{\rm i} (\rho_1-\rho_4)},
 \label{Eq:Heat}
 \end{equation}
where we denoted $H_{\rm i}=H(\lambda_{\rm i})$, $H_{\rm f}=H(\lambda_{\rm f})$, $\rho_\alpha$ for $\alpha=\{1,2,3,4 \}$ represents the density matrix in the corresponding stages of the cycle, cf. Fig.~\ref{fig1}.
Formulas in Eqs.~\eqref{Eq:Heat} are constructed in the way that if the heat is transfered into the WF then $Q>0$ and if transfered out then $Q<0$.
As no heat is transfered between the WF and the reservoirs in the strokes  $1 \to 2$ and $3 \to 4$, the First law of thermodynamics gives the amount of work per closed cycle as $W=-(Q_h+Q_c)$.
This formula can be used for the finite-time operational mode of the engine as well, provided that the limit cycle has been already reached.
Note that in our convention the extracted work has negative sign $W<0$ (let us denote this quantity simply as $W'=-W$).
The efficiency of the engine $\eta=W'/Q_h$ is bounded by the Carnot efficiency $\eta_C=1-T_c/T_h$.

\subsection{Reference temperature}
\label{Subsec:RefTemp}
For any state of the system described by a density matrix $\rho$ and Hamiltonian $H$ we can define a \textit{reference} thermal state~\cite{Ali13,Gel19} $\rho^*$ by equating the entropies $S(\rho)=-\trace{\rho \ln{\rho}}$ and imposing the Gibbs form of $\rho^*$
\begin{equation}
\rho^*=\frac{e^{-\beta^*H}}{\trace{e^{-\beta^*H}}}, \qquad S(\rho)=S(\rho^*).
\label{Eq:VirtTempDef}
\end{equation}
Quantity $\beta^*$ is referred to as an inverse reference temperature (we define also the reference temperature $T^*=1/\beta^*$).
When restricted to positive values only $\beta^*>0$ it can be uniquely assigned to any state via Eq.~\eqref{Eq:VirtTempDef}.
If one thinks of entropy from the viewpoint of information theory, the corresponding thermal reference state $\rho^*$ minimizes the energy while keeping the same amount of information.
The energy difference between the actual state and the thermal reference state is used to set the upper bound on \textit{ergotropy}, i.e., the maximal extractable work with unitary transformations~\cite{Ali13}. 


Generally, the reference temperature is not a real temperature of the system which may be in an arbitrary nonequilibrium state.
One example where it is, however, so is the case of a qubit.
Indeed, any diagonal qubit state with decreasing populations as a function of energy can be considered as thermal.
In other cases the reference temperature can still provide some intuitive insight.
When the Hamiltonian $H$ is constant in time, the state $\rho$ may still change either due to unitary evolution, Lindblad evolution, etc.
Then, the inverse intrinsic temperature $\beta^*$ becomes time-dependent.
One can show that
\begin{equation}
\frac{dS}{dt} = -\frac{d\beta^*}{dt} \beta^*(t) \left[\ave{H^2}_* - \ave{H}_*^2\right]\,. 
\end{equation}
Since the expression in brackets is always positive, it means that the entropy increases when the reference temperature increases and vice versa.

More specifically, for a Davies-Lindblad map -- microscopically implementing a thermal reservoir at inverse temperature $\beta$ -- one can express the change of the system entropy also as~\cite{Dav74}
\begin{equation}
\frac{dS}{dt} = \beta \dot{Q} -\trace{({\cal L}\rho) \left[\ln \rho - \ln \rho_\beta\right]}\,,
\label{Eq:heatflow}
\end{equation}
where $\rho_\beta$ just denotes the thermal Gibbs state with temperature $\beta$ and  ${\cal L}$ is the evolution superoperator with ${\cal L} \rho_\beta = 0$.
In particular, the second term is always positive due to Spohn's inequality $-\trace{({\cal L}\rho) \left[\ln \rho - \ln \rho_\beta\right]} \ge 0$ (see Refs.~\cite{Ali79,Spo78} and references therein).
The quantity $\dot{Q}$ denotes the heat current entering the system from the reservoir.

From this, we can conclude \textit{i)} if the heat current is positive, the reference temperature must increase, \textit{ii)} if the reference temperature decreases, the heat current must be negative.
However, we cannot infer the corresponding opposite, i.e., an increasing reference temperature does not imply that the heat current is positive.

%

\subsection{Fidelity}
There exist several measures on the space of density matrices which quantify the distance between individual states.
In the present work we use fidelity which takes two density matrices  $\rho$ and $\sigma$ as arguments~\cite{Nie00}
\begin{equation}
 \mathcal{F}(\rho||\sigma)= \trace{\sqrt{\sqrt{\rho}\sigma\sqrt{\rho}}}^2\,.
\label{Eq:Fidelity}
\end{equation}
It will be used below to monitor the departure of the actual state of the WF from the thermal reference state.

In principle, any other measure could be used with the same qualitative results.
The reasons why we favor fidelity is that it is symmetric $\mathcal{F}(\rho||\sigma)=\mathcal{F}(\sigma||\rho)$ and bounded $0 \leq \mathcal{F}(\rho||\sigma) \leq 1$ where the maximum is achieved for $\rho=\sigma$.
Moreover, for pure states $\rho=\ket{\phi_\rho}\bra{\phi_\rho}, \ \sigma=\ket{\phi_\sigma}\bra{\phi_\sigma}$ it reduces to the simple form $\mathcal{F}(\rho||\sigma)= \abs{\scal{\phi_\rho}{\phi_\sigma}}^2$ having a direct intuitive meaning.

\section{Non-interacting spin model}
\label{Sec:NonInt}

Due to the absence of interaction, we  neglect in this Section all effects of coherences in the eigenbasis of the WF Hamiltonian.
They would decay during thermalization anyway and are not restored during the unitary strokes.
Therefore the only genuinely quantum feature in this Section is the discrete spectrum of the WF.

\subsection{Single qubit}
\label{Sec:SingleQbit}

We start by considering the WF composed of mutually non-interacting spins (qubits), first treating a single one only
\begin{equation}
H(t)=-\frac{\lambda(t)}{2}\omega \sigma_z\,,
\label{Eq:SingleQBIT}
\end{equation}
where $\lambda(t)$ is a time-dependent dimensionless parameter and $\sigma_z$ is the Pauli matrix.
Parameter $\omega$ defines the energy scale of the model.

This model was used to set benchmark conditions on the performance of the Otto cycle~\cite{Kie04,Kie06}.
As there is no interaction in unitary strokes $1\to 2$ and $3\to 4$, only the energy gap between the levels is altered and the process is inherently quantum adiabatic (classically, \lq adiabatic\rq\ just means no heat exchange, so any unitary stroke would always be adiabatic).
For the same reason the density matrix stays unchanged in these strokes, i.e., $\rho_1=\rho_2$ and $\rho_3=\rho_4$.
Using Eqs.~\eqref{Eq:Heat}, the efficiency can be expressed simply as 
\begin{equation}
\eta=1-\frac{\lambda_{\rm i}}{\lambda_{\rm f}},
\label{Eq:Kieu}
\end{equation}
 where $\lambda_{\rm i}$ and $\lambda_{\rm f}$ are the initial and final values of the parameter $\lambda$ respectively (assuming $\lambda_{\rm i}<\lambda_{\rm f}$).

The master equation to model the thermalization strokes can be written as follows~\cite{Bre02,Wei12}
\begin{equation}
  \dot{\rho}=\ii\frac{\lambda}{2}\omega[\sigma_z,\rho]+\gamma(1+n_b)\mathcal{D}[\sigma_+]\rho+\gamma n_b \mathcal{D}[\sigma_-]\rho,
  \label{Eq:Thermalization}
 \end{equation}
 with the Lindblad dissipators $\mathcal{D}[O]\rho=O\rho O^\dagger-\frac{1}{2} \{O^\dagger O,\rho  \}$. 
The temperature of the heat bath $\beta=1/T$ with $T\in\{T_c,T_h\}$ is contained in the Bose-Einstein  distribution factor $n_b=(e^{\beta \lambda}-1)^{-1}$, $\gamma$ denotes the dissipation rate.
Such a Lindblad equation arises from the subsequent application of Born-Markov and secular approximations~\cite{Bre02}, we therefore expect it to be valid in the regimes $\gamma\beta \ll 1$ and $\lambda \omega \gg \gamma$.

\begin{figure}[t]
	\centering
  \includegraphics[width=1\linewidth, angle=0]{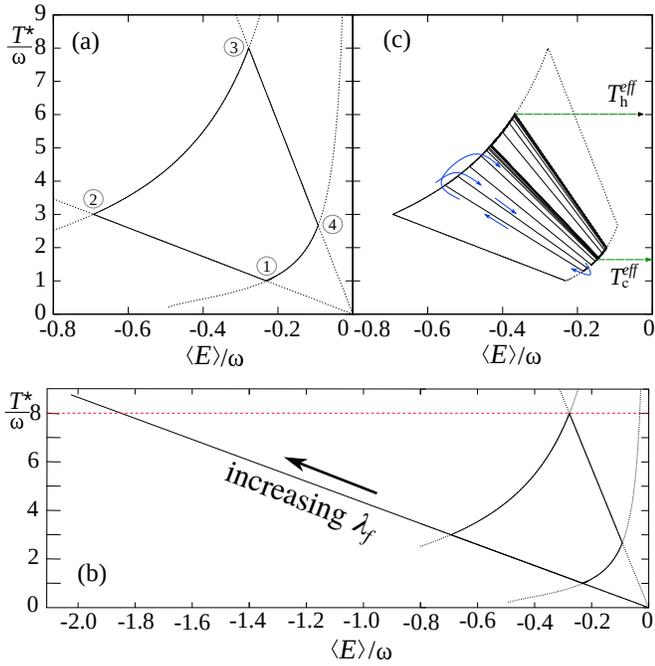}
	\caption{Single qubit in Otto cycle with parameters $\lambda_{\rm i}=1,\  \lambda_{\rm f}=3, T_c=1\omega, \ T_h=8 \omega, \ \gamma=0.1 \omega$.
	Panel~(a): Cycle with a full thermalization. The numbers in circles correspond to Fig.~\ref{fig1}. The dotted curves are analytic.
	Panel~(b): A visual demonstration of the Carnot bound. 
	For description see the main text.
	Panel~(c): System approaching the limit cycle in the case of a finite-time thermalization. Duration of the contact with the heat baths is fixed as~$t_{\rm th}=1 \omega^{-1}$.
	The arrows show how the limit cycle is reached by \lq winding\rq\ around the full thermalization cycle (plotted with dotted curves).
	}
	\label{fig2}
\end{figure}

We will study the evolution of the system in the plane $\langle E \rangle \times T^*$ where  $\ave{E}$ is the mean energy and $T^*=1/{\beta^*}$ is the reference temperature.
We stress again that in a two-level system any diagonal state with decreasing populations with energy can  be considered as thermal so the reference temperature is directly linked with the thermodynamic one.
In Fig.~\ref{fig2}(a) the full Otto cycle with a single qubit is depicted.
The cycle can be reconstructed analytically.

The unitary parts $1\to 2$ and $3\to 4$ are unavoidably quantum adiabatic and they show linear dependence between $T^*$ and $\langle E \rangle$.
The occupation probabilities for the excited $p_{\rm e}$ and the ground states $p_{\rm g}$ remain constant.
Suppose mean energy $\ave{E}_0$ and the corresponding temperature $T_0$ are known for a certain value $\lambda_0$.
We can then write
\begin{equation}
\ave{E}_\lambda = -\frac{\lambda}{2}\omega\trace{\sigma_z \rho} \Rightarrow \frac{\ave{E}_\lambda}{\ave{E}_{0}}=\frac{\lambda}{\lambda_0}\,,
\label{Eq:MeanE}
\end{equation}
\begin{equation}
p_{\rm e} \propto e^{{-\frac{\lambda_0 \omega}{2T_0}}}=const. \Rightarrow T^*_\lambda=\frac{\lambda}{\lambda_0}T_0\,.
\label{Eq:Temp}
\end{equation}
Combining Eqs.~\eqref{Eq:MeanE} and~\eqref{Eq:Temp} we obtain
\begin{equation}
 T^*= \frac{T_0}{\ave{E}_0}\ave{E}.
\label{Eq:LinearAdiabtic}
\end{equation}
For example, in position $1$ of the cycle the system is at temperature $T_0=T_c$.
The corresponding mean energy can be computed $\ave{E}_0 \approx -0.231 \omega$.
From Eq.~\eqref{Eq:LinearAdiabtic} we obtain the evolution in Fig.~\ref{fig2}(a) between $1 \to 2$.
Similarly the evolution between $3 \to 4$ can be obtained by considering $T_0=T_h$ (at point 3) and $\ave{E}_0\approx-0.278\omega$.

Thermalization in strokes $2 \to 3$ and $4\to 1$ is performed with $\lambda$ fixed.
Along the thermalization process we always write the occupation probability of the excited state in the form of a thermal state
\begin{equation}
p_{\rm e}=\frac{e^{-\frac{\lambda \omega}{2T^*}}}{Z(T^*)}=\frac{1}{1+e^{\frac{\lambda \omega}{T^*}}}\,,
\label{Eq:Prob}
\end{equation}
where $Z(T^*)$ is the partition sum.
Considering $1=p_{\rm e}+p_{\rm g}$, the mean energy can be expressed as $\ave{E}/\omega =\frac{\lambda}{2}(2p_{\rm e}-1)$.
Combining this with Eq.~\eqref{Eq:Prob} we can express 
\begin{equation}
 T^*=\frac{\lambda\omega}{\ln{\Big(\frac{\lambda\omega-2\ave{E}}{\lambda\omega+2\ave{E}}\Big)}}.
\label{Eq:NonLinTherm}
\end{equation}

The map $ \ave{E} \times T^* $  can provide some insight simply based on visual inspection.
As we change parameter $\lambda$ in the unitary strokes we effectively \lq heat up\rq\ or \lq cool down\rq\ the system in a linear way.
For example in the stroke $1 \to 2$, there always exists a point where this linear dependence reaches the temperature of the heat bath $T_h$, see Fig~\ref{fig2}(b).
If by changing $\lambda_{\rm f}$ this point is crossed then the machine cannot work as a heat engine because no heat is transfered from the heat reservoir in the subsequent stroke.
According to Eq.~\eqref{Eq:Temp} the relation between the initial and final temperature in the stroke is $T^*_{\rm f}=\frac{\lambda_{\rm f}}{\lambda_{\rm i}}T^*_{\rm i}$.
Considering we start from the thermal equilibrium state of the cold reservoir  $T_{\rm i}=T_c$, positive work can be extracted in the cycle only if $T_{\rm f}<T_h$.
We obtain the condition $\lambda_{\rm f}/\lambda_{\rm i}<T_h/T_c$ which guarantees that the efficiency given by Eq.~\eqref{Eq:Kieu} is bounded by Carnot's value $\eta_C$.

%

Now we prepare the WF in a thermal equilibrium with the cold bath at $T_c$ and evolve it in a way that in the thermalization segments of the cycle it will be in contact with the heat bath for only $t_{\rm th}=1 \omega^{-1}$.
During this time the WF is unable to fully thermalize, see Fig.~\ref{fig2}(b).
After a few cycles the system reaches a stable operational mode represented by a limit cycle in the plane $\ave{E} \times T^*$ which is approached by \lq winding\rq\ around the full thermalization cycle.
The reason that the evolution does not deviate from it is grounded in the fact that during the cycle the population distribution in the qubit WF stays precisely thermal.
So the unitary evolution always oscillates between the thermalization curves given by Eq.~\eqref{Eq:NonLinTherm} for $\lambda_{\rm i}$ and $\lambda_{\rm f}$.
As the reference temperature coincides with the thermodynamic one for a two-level system, the stable mode of operation is equivalent to a fully thermalized Otto cycle working between different \textit{effective} heat baths.
Their temperatures $T_c^{\rm eff}, \ T_h^{\rm eff}$ can be identified as the lowest and highest points of the limit cycle, respectively.
The efficiency of such a machine is still given by Eq.~\eqref{Eq:Kieu} and is bounded by Carnot value given by the real bath temperatures $T_c, \ T_h$, so it does not differ from the fully thermalized regime.
The work extracted in a cycle is smaller but can be gained faster compared to the case when one operates the machine between the real heat baths with temperatures $T_c^{\rm eff}$ and $\ T_h^{\rm eff}$.
So the finite-time machine can outperform the one with fully thermalized strokes in terms of the power output as will be explicitly shown later.


\subsection{Collective spin model, superradiant effect}
\label{SubSecCollective}

In this Section we consider the Hamiltonian of $N=2j$ copies of a single qubit written using collective spin operators
 \begin{equation}
 H(t)=-\lambda(t) \omega J_z, \quad \ J_\alpha=\sum_{i=1}^{2j}\frac{\sigma_\alpha^{(i)}}{2}, \quad \alpha={x,y,z}\,.
 \label{Eq:Ham}
 \end{equation}
 If they thermalize incoherently (without any mutual correlations) then the previous Section is applicable as there are $N$ independent qubits forming the WF.
 Here we consider coherent dissipation~\cite{Aga74,Vog11} with $J_\pm=J_x \pm \ii J_y$ according to the equation
 \begin{equation}
  \dot{\rho}=\ii\lambda\omega[J_z,\rho]+\gamma(1+n_b)\mathcal{D}[J_+]\rho+\gamma n_b \mathcal{D}[J_-]\rho.
  \label{Eq:CohDis}
 \end{equation}

\begin{figure}[t!]
	\centering
  \includegraphics[width=1\linewidth, angle=0]{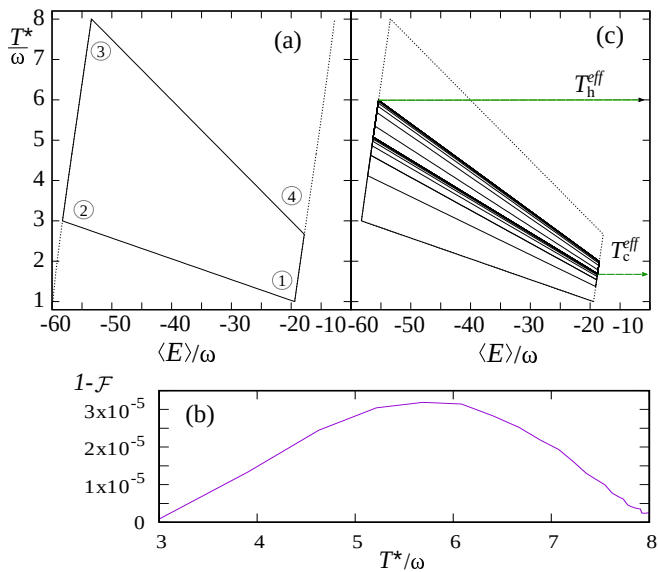}
	\caption{Collective spin system~\eqref{Eq:Ham} in Otto cycle with parameters $j=20,\ \ \lambda_{\rm i}=1,\   \lambda_{\rm f}=3,\   T_c=1\omega, \  T_h=8\omega, \  \gamma=0.1 \omega$.
	Panel~(a): Cycle with a complete thermalization. The numbers in circles correspond to Fig.~\ref{fig1}. The dotted curves represent truly thermal states for $\lambda_{\rm i}$ and $\lambda_{\rm f}$.
	Panel~(b): Distance between the actual state $\rho$ and the thermal reference state $\rho^*$ during the thermalization stroke $2 \to 3$ measured by fidelity $\mathcal{F}$.
	Panel~(c): System approaching the limit cycle in the case of a finite-time thermalization. Duration of the contact with the heat baths is fixed as~$t_{\rm th}=0.1 \omega^{-1}$.
	The dotted cycle represents the full thermalization case.
	}
	\label{fig3}
\end{figure}

In Fig.~\ref{fig3}(a) we show the full Otto cycle with the system of the size $j=20$.
In many aspects it behaves similarly to the single qubit case.
Again, the density matrix does not change during the unitary strokes and the changes of $\lambda$  only uniformly modify the gaps between the levels.
If originally the WF was in the thermal state then varying $\lambda$ effectively \lq heats up\rq\ or \lq cools down\rq\ the WF.
The reason is simply that the distribution of occupation probabilities is only uniformly stretched or shrunk and so keeps its thermal nature.
This means that Eq.~\eqref{Eq:LinearAdiabtic} for quantum adiabatic strokes is still valid and the reference temperature coincides with the thermodynamic one.
In the same way as discussed in the previous Sec.~\ref{Sec:SingleQbit} one can conclude that the efficiency is still given by Eq.~\eqref{Eq:Kieu} and bounded by the Carnot value.

During the thermalization parts of Fig.~\ref{fig3}(a) it is, however, not  guaranteed that the system passes through truly thermal states so $T^*$ cannot be generally associated with thermodynamic temperature.
As can be numerically verified, during the stroke the WF deviates from the thermal state with  $T^*$ but it still stays remarkably close to it.
This is visible in  Fig.~\ref{fig3}(b) where the distance between the actual state $\rho$ and the reference thermal state $\rho^*$ expressed via fidelity~\eqref{Eq:Fidelity} is plotted.
One can see that the maximal deviation is of order $10^{-5}$ so the reference thermal state approximates the real state very well.
 
The evolution of the finite-time heat engine with $t_{\rm th}=0.1 \omega^{-1}$ is depicted in  Fig.~\ref{fig3}(c).
The limit cycle is formed in a similar way as in a single qubit case.
However, now the equivalence to a fully thermalized cycle with two effective heat baths is only approximate because of the arguments in the paragraph above. 

\begin{figure}[t!]
	\centering
  \includegraphics[width=0.9\linewidth, angle=0]{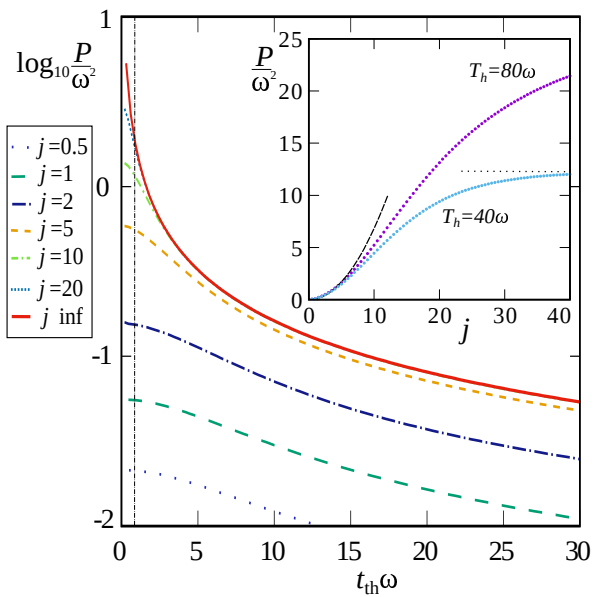}
	\caption{Power of the collective-spin heat engine as a function of the duration of the thermal strokes $t_{\rm th}$.
	The curve for $j\to \infty$ is analytic according to Eq.~\eqref{Eq:WcLargeJ}.
	Parameters are $T_c=1 \omega, \ T_h=8 \omega, \ \lambda_{\rm i}=1, \ \lambda_{\rm f}=3, \ \gamma=0.1 \omega$.
	Inset: Power as a function of the size of the system $j$ computed for fixed $t_{\rm th}=1 \omega^{-1}$ (denoted with a thin vertical line in the main part) and two different values of $T_h$ as indicated (other parameters are the same as in the main part of the figure). The dashed curve represents the quadratic fit for $j\le 5$. The dotted line indicates the saturation value of power  $\overline{P}\approx 12.26 \omega^2$  [given by Eq.~\eqref{Eq:WcLargeJ}] for the $T_h=40\omega$ bath. The maximal power for the $T_h=80\omega$ bath is $\overline{P}\approx 25.59 \omega^2$.
	}
	\label{fig4}
\end{figure}

As has been already pointed out, for any $j$ the efficiency is still the same regardless of the operational mode (fully-thermalized vs. finite-time).
Work extracted in the limit cycle is apparently decreasing by making $t_{\rm th}$ smaller.
However, the system can run through the limit cycle very quickly, so the power of the machine in this setting can overcome the mode with full thermalization.
Indeed, in Fig.~\ref{fig4} we present the dependence of power $P$ as a function of the duration of the thermal strokes $t_{\rm th}$. 
Power is computed as $P=W'_c/t_c$ where $W'_c$ is the work output in the limit cycle and $t_c$ is its duration.
As in the current setting the unitary strokes can be arbitrarily fast, we simply put $t_c=2 t_{\rm th}$.

Generally, for smaller values $t_{\rm th}$ we can get higher power from the system regardless of the size $j$.
For any  $j$ the power per limit cycle is a monotonously decreasing function of $t_{\rm th}$.
There also exists a certain value $t_{\rm th}=t_T$ (thermalization time) where the WF can be considered as fully thermalized and so by further enlarging   $t_{\rm th}$, one does not extract any more work.
As a result, for $t_{\rm th}>t_T$ the power must behave simply as $\propto 1/t_{\rm th}$.

The maximal power output could be naively extracted by taking $t_{\rm th} \to 0$ but this limit is singular (for $t_{\rm th}=0$ there is no contact with the baths so the work output is zero) and, of course, practical realization of \textit{very small values} of $t_{\rm th}$ is limited.
Nevertheless, the results in Fig.~\ref{fig4} show that operating the engine in the regime $t_{\rm th}<t_T$ is beneficial for the power output.

Now we turn our attention to the performance of the heat engine as a function of $j$.
We express Eq.~\eqref{Eq:CohDis} in the eigenbasis  $J_z\ket{m}=m\ket{m}$ and focus on the dynamics of the diagonal terms $\rho_m \equiv \scal{m|\rho}{m}$ (the coherences evolve independently and decay during the thermalization)
\begin{equation}
\begin{split}
\dot{\rho}_m=&\gamma (1+n_b)(j+m)(j-m+1)\rho_{m-1}\\
&+\gamma n_b(j-m)(j+m+1)\rho_{m+1}\\
&-\gamma(1+n_b)[j(j+1)-m(m+1)]\rho_m\\
&-\gamma n_b [j(j+1)-m(m-1)]\rho_m.
\end{split}
\label{Eq:diagonalsEvo}
\end{equation}
The Clebsch-Gordon coefficients in front of the terms $\rho_{m\pm 1}$ are of order $j$ at both edges of the spectrum $\ave{J_z}\approx \pm j$ whereas  in the central region $\ave{J_z}\approx 0$ they scale as $j^2$.
The latter coefficients are responsible for the well-known superradiant relaxation at zero temperature~\cite{Dic54,Gro82}.
Our case generalizes the situation to the finite-temperature regime, nevertheless, due to the large Clebsch-Gordon coefficients we can still expect some superradiant $N^2$ (or $j^2$) scaling of the engine when eigenstates with $m \approx 0$ are populated. 

The inset of Fig.~\ref{fig4} shows that this scaling can appear in power output of the machine operated at the fixed time $t_{\rm th}$, however in a rather small domain of values $j$.
The reason is that in order to observe the superradiant enhancement, at least one of the thermal reservoirs must have sufficiently large temperature so that the  $\ave{J_z}\approx 0$ region becomes populated (in the inset of Fig~\ref{fig4} we consider $T_h=40 \omega$ and $T_h=80 \omega$).
Obviously, for growing $j$ one would need higher and higher temperatures to keep this region occupied.
So the initial quadratic scaling, representing a superradiant boost in power, reduces to the linear with growing $j$, unless the hot reservoir is kept at infinite temperature.

The inset of Fig.~\ref{fig4} also shows that for $j\to\infty$ the power output saturates at the maximal value $\overline{P}$.
Maximal power $\overline{P}$ as a function of $t_{\rm th}$  can be computed analytically and is shown in the main part of Fig.~\ref{fig4}. 
The saturation results from the maximal extractable work  harvested in the  large-$j$ limit $\overline{W}'_c=\lim_{j\to\infty}W'_c$ which is finite.
Using Eqs.~\eqref{Eq:Heat} and the fact that in our setting $\rho_1=\rho_2$ and $\rho_3=\rho_4$ (as was already pointed out) we obtain
 \begin{equation}
 \overline{P}=\frac{\overline{W}'_c}{t_c}=\frac{\Delta \lambda \omega}{2t_{\rm th}}\frac{e^{\beta_c\lambda_{\rm i}\omega}-e^{\beta_h\lambda_{\rm f}\omega}}{(e^{\beta_h\lambda_{\rm f}\omega}-1)(e^{\beta_c\lambda_{\rm i}\omega}-1)}\,,
 \label{Eq:WcLargeJ}
 \end{equation}
 where $\Delta \lambda= \lambda_{\rm f}-\lambda_{\rm i}$.
 The same result would be obtained for an adiabatically driven harmonic oscillator with frequency $\lambda(t) \omega$~\cite{Kos17}.

If one inserts the values of $\lambda_{\rm i}=T_c/\omega$ and $\lambda_{\rm f}=T_h/\omega$ so that the Carnot maximal efficiency~\eqref{Eq:Kieu} is achieved, then Eq.~\eqref{Eq:WcLargeJ} gives zero power output, as expected.
The efficiency at the maximum power is well approximated by the Curzon-Ahlborn (Chambadal-Novikov) value~\cite{Nov54,Cha57,Cur75}
\begin{equation}
\eta_{CA}=1-\sqrt{\frac{T_c}{T_h}}\,,
\end{equation}
which is reached by setting parameters $\lambda_{\rm i}=\sqrt{T_c/\omega}$ and $\lambda_{\rm f}=\sqrt{T_h/\omega}$.
 The corresponding performance is then
 \begin{equation}
 \overline{P}_{CA}=(\sqrt{T_h}-\sqrt{T_c})\frac{\sqrt{\omega}}{2t_{\rm th}}\frac{e^{\frac{\omega}{T_c}}-e^{\frac{\omega}{T_h}}}{(e^{\frac{\omega}{T_h}}-1)(e^{\frac{\omega}{T_c}}-1)}\,.
 \end{equation} 
 More precise analytic estimations on the efficiency at maximal power which employ the same or similar systems can be found in Refs.~\cite{Sch08,Wan13,Wu14,Dor18,Abi19}.

Let us compare the current situation to that of independent qubits with incoherent dissipation.
We already showed that there exists a region where the power is boosted as $N^2$  compared to the incoherent case where one simply gets the linear scaling.
The growth of power is bounded by $\overline{P}$ from Eq.~\eqref{Eq:WcLargeJ} and so the quadratic dependence occurs only when the machine is operated at $t_{\rm th} < t_T$.

Now, we focus on the regime $t_{\rm th} \approx t_T$.
Work extracted in a fully thermalized cycle for a single qubit is
 \begin{equation}
 W_{\rm qb}'=\Delta \lambda \omega\frac{e^{\beta_c\lambda_{\rm i}\omega}-e^{\beta_h\lambda_{\rm f}\omega}}{(e^{\beta_h\lambda_{\rm f}\omega}+1)(e^{\beta_c\lambda_{\rm i}\omega}+1)}\,.
 \label{Eq:WsingleQB}
 \end{equation}
Obviously, for $N$ such qubits we gain work of the total amount $N W_{\rm qb}'$ which goes to infinity with $N=2j\to\infty$.
On the other hand when these qubits dissipate coherently their work output is finite in $j\to\infty$ as shown in Eq.~\eqref{Eq:WcLargeJ}.
Does that imply that the large sample of incoherently dissipating qubits should now be  favored  in terms of power?
Not really.
In reality, the opposite statement is true.

\begin{figure}[t!]
	\centering
  \includegraphics[width=1\linewidth, angle=0]{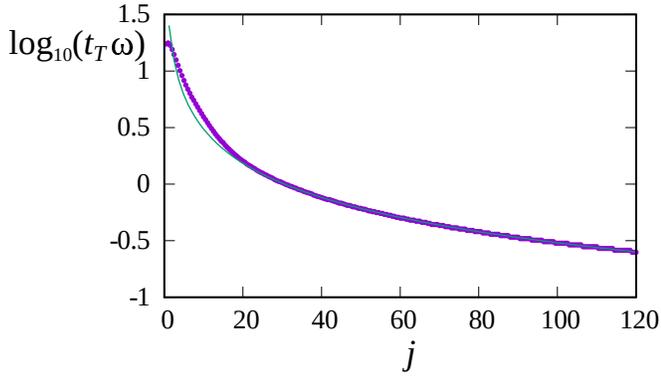}
	\caption{Thermalization time $t_T$ as a function of $j$ according to Eq.~\eqref{Eq:CohDis}.
	Fidelity $\mathcal{F}(\rho ||\rho_{T_{\rm f}})$ was used as a measure of the distance between the actual state $\rho$ and the final thermal state $\rho_{T_{\rm f}}$, see Eq.~\eqref{Eq:Fidelity}. The case shown corresponds to cooling of the thermal state from $T_{\rm i}=4 \omega$ to $T_{\rm f}=1 \omega$ with $\lambda=1$ fixed, $\gamma=0.1 \omega$.
	The tolerance to establish $t_T$ was chosen as $1-\mathcal{F}\le 10^{-5}$.
	The green curve is a $1/j$ fit.
	}
	\label{fig5}
\end{figure}

The key observation is that $W_{\rm qb}'$ as well as $\overline{W}'_c$ are reached under the condition of a \textit{fully} thermalized cycle and so a relevant comparison of the power output must be made for the precise corresponding thermalization times $t_T$.
Fig.~\ref{fig5} shows the dependence of $t_T$ on the size of the system $j$ for the coherent case.
We observe that $t_T$ decreases as~$1/j\sim1/N$.
 In contrast, for the incoherent case, essentially, the thermalization time corresponds to the one of a single qubit $t_T^{\rm qb}$ regardless of the size of the ensemble.
 
So an optimal setting to harvest work $W_{\rm qb}'$ or $\overline{W}'_c$ is to operate the machine with the corresponding $t_T^{\rm qb}$ or $t_T(N)$.
We define a relative power output at these optimal times for the large system limit as
\begin{equation}
\mathcal{P}=\lim_{N \to\infty}\frac{\overline{P}(t_T(N))}{N P_{\rm qb}(t_T^{\rm qb})}\,,
\label{Eq:RelP}
\end{equation} 
where $P_{\rm qb}=W_{\rm qb}'/t_c$ is a single qubit performance in the cycle.
Due to the dependence $t_T(N)=\alpha/N$ the limit in Eq.~\eqref{Eq:RelP} is non-zero.
Apparently, the constant $\alpha$ can be read off as the value $t_T(N=1)$.
As the fitting function in Fig.~\ref{fig5} corresponds to the numerical data well even in the region of small $j$, we approximate $\alpha$ by the real single qubit thermalization time $t_T^{\rm qb}$.
Thus we obtain
\begin{equation}
\mathcal{P}\approx \coth{\left(\frac{\beta_h\lambda_{\rm f}\omega}{2}\right)}\coth{\left(\frac{\beta_c\lambda_{\rm i}\omega}{2}\right)>1}\,,
\label{Eq:RepPII}
\end{equation}
showing that the power output is larger in the case of coherent dissipation.

Now it is clear that the coherent dissipation is beneficial in terms of a power gain both in the region $t_{\rm th}<t_T$ and $t_{\rm th}\approx t_T$ (region $t_{\rm th}> t_T$ is generally unfavorable as no further work is extracted).
As already mentioned, this cooperative boost in power represents a close analog of the Dicke superradiance phenomenon, i.e., the collective enhancement of coherent spontaneous emission from a dense ensemble of atoms~\cite{Nie18b,Dic54, Gro82}.
In the original setting the atoms interact with each another through a common radiation field.
In analogy to that the interaction among the qubits in the current case is mediated by a common heat bath and the collective dissipators.

\begin{figure}[t!]
	\centering
  \includegraphics[width=0.85\linewidth, angle=0]{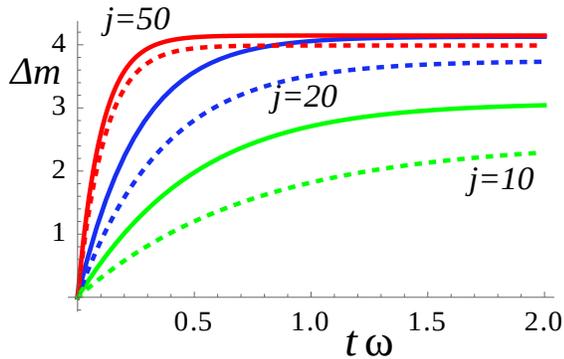}
	\caption{Analytic solution to the thermalization from a thermal state of $T_{\rm i}=8\omega$ to $T_{\rm f}=4\omega$ using Eq.~\eqref{Eq:AnalogSuperrad} for three different values of $j$. We define $\Delta m(t) = m(t)-m_0$. Value $\lambda=1$ is fixed, $\gamma=0.1 \omega$. The dotted lines are the numerical solutions.
	}
	\label{fig6}
\end{figure}

Pushing this analogy forward, motivated by Ref.~\cite{Gro82} we derive the equation for the time evolution of the expectation value of $J_z$ using Eq.~\eqref{Eq:CohDis}
 \begin{equation}
  \dot{\ave{J_z}}=-\gamma(1+2n_{b})\ave{J_z}-\gamma\ave{J_z^2}+\gamma j(j+1)\,.
 \label{Eq:AnalogSuperrad}
 \end{equation}
Applying the mean-field approximation $\ave{J_z^2} \approx \ave{J_z}^2$ the equation can be solved analytically.
We further denote $m(t) \equiv \ave{J_z}$ and consider it to be continuous.
The mean-field solution to Eq.~\eqref{Eq:AnalogSuperrad} is
   \begin{eqnarray}   
   m(t)&=&-\frac{1}{2}(1+2n_{b})+C \tanh{\big(C\gamma(t-\tilde{t})\big)}, \label{Eq:SuperRadAnalog}
   \\
   C&=&\frac{1}{2}\sqrt{4j(j+1)+(1+2n_{b})^2}. \nonumber
    \end{eqnarray}
Equation~\eqref{Eq:SuperRadAnalog} gives qualitatively the same type of time dependence $m(t)$ as in the standard superradiant setting~\cite{Gro82}.
Time shift $\tilde{t}$ is determined by the initial value of $m_0 \equiv m(0)$.
If the initial state is thermal with initial inverse temperature $\beta_{\rm i}$ then
\begin{eqnarray}
 m_0&=& \frac{\trace{J_z e^{-\beta_{\rm i}H}}}{\trace{e^{-\beta_{\rm i}H}}}\,, 
 \label{Eq:m0}
\end{eqnarray}
The value $\tilde{t}$ can then be expressed (using exponential expansion of hyperbolicus tangent)
\begin{equation}
\tilde{t}= \frac{1}{2C\gamma}\ln{\Big(\frac{-1+2C-2m_0-2n_{b}}{1+2C+2m_0+2n_{b}}\Big)}.
 \label{Eq:t0}
\end{equation}
It is negative and converges to $0$ for $j\to\infty$.
This is different from the standard superradiance where $\tilde{t}$ would  define the time of the superradiant burst (so apparently its value must be positive).
However, qualitatively the solution is the same in this case and explains the speed-up in thermalization.
This is demonstrated in Fig.~\ref{fig6} where the simple analytic solution given by Eq.~\eqref{Eq:SuperRadAnalog} shows how the steady state given by $\Delta m(t)= m(t)-m_0 = const.$ is reached faster for growing $j$.
The analytic formula is also compared with the numerical results.
We see the improvement of the mean-field approximation as $j$ becomes larger.

The  solution~\eqref{Eq:SuperRadAnalog} can as well be used to show the $1/j$ dependence in the thermalization time $t_T$ for large $j$ as depicted in Fig.~\ref{fig5}.
From Eq.~\eqref{Eq:SuperRadAnalog} one obtains the analytic approximation of the steady state considering limit $t\to\infty$
\begin{equation}
m_{\rm ss}=C-\frac{1}{2}(1+2 n_b)\,.
\label{Eq:ApSteadyState}
\end{equation}
We can represent the \lq fidelity\rq\ as the distance
\begin{equation}
{\rm dist}(t)=\frac{\abs{m_{\rm ss}-m(t)}}{m_{\rm ss}}\,,
\label{Eq:ApDist}
\end{equation}
and we can introduce a condition that we consider the system to be thermal if ${\rm dist}(t) < \varepsilon$ where $\varepsilon$ defines the precision.
Considering $j\gg 1$ and setting $\tilde{t}=0$ for the sake of simplicity, one obtains the condition for $t_T$
\begin{equation}
1-\tanh{(j\gamma t_T)}=\varepsilon\,,
\label{Eq:AptT}
\end{equation}
from which the dependence $t_T \propto 1/j$ is clear.

\section{Lipkin-Meshkov-Glick model}
\label{Sec:LMG}
In this Section we turn to the situation when the spins forming the WF mutually interact. 
We consider a collective long-range interaction of the the Lipkin-Meshkov-Glick (LMG) type~\cite{Lip65}.
Due to the interaction the unitary evolution is generally quantum non-adiabatic, unless \textit{sufficiently} slow.
For the same reason the coherences are built up in the basis of $J_z$.

\subsection{The protocol}
The LMG Hamiltonian is taken in the form
\begin{equation}
H(t)=-\lambda(t)\omega J_z - \frac{\Gamma(t)\omega}{N}J_x^2, \ \ N=2j\,. 
\label{Eq:LMG}
\end{equation}
The Hamiltonian is time-dependent through the dimensionless control parameters $\lambda(t)$ and $\Gamma(t)$.
Similarly to the previous Section, parameter $\omega$ sets the energy scale of the system.
Hamiltonian \eqref{Eq:LMG} conserves parity $\Pi=e^{\ii \pi (J_z+j)}$ and so the states from different parity sectors do not interact.

As we want to be able to model the thermalization strokes with Eq.~\eqref{Eq:CohDis} we have to guarantee that during these segments of the cycle $\Gamma(t)=0$.
So we consider the following protocol for varying the parameters during the unitary strokes
\begin{eqnarray}
\lambda(t)&=&\lambda_{\rm i}[1-s(t)]+\lambda_{\rm f} s(t), \label{Eq:prot1}
 \\ \Gamma(t)&=&4 \bar{\Gamma} s(t) [1-s(t)]\,. \label{Eq:prot2}
 \end{eqnarray}
The function which inserts the time dependence is a simple linear ramp $s(t)=t/t_u$ where $t_u$ defines the overall duration of a unitary stroke.
One can easily check that for $t=0$ and  $t=t_u$ the system is described by a non-interacting Hamiltonian~\eqref{Eq:Ham}.
Constant parameter $\bar{\Gamma}$ defines the maximal value of $\Gamma (t)$ reached during the stroke.

In our protocol the LMG model is coupled to the thermal baths with $\Gamma(t)=0$ and we focus on the system with large $j$.
Therefore, the findings of the previous Section on collective equilibration are directly applicable.
In the following we always consider full thermalization in the cycle (again, within a given tolerance) which is achieved in a short thermalization time $t_T$ due to the superradiant effect. 
Therefore, we focus solely on the effects of finite-time unitary strokes as they are crucial for the work and power output of the engine in this setting.

\subsection{Criticality and the reference temperature}
\begin{figure}[t!]
	\centering
  \includegraphics[width=1\linewidth, angle=0]{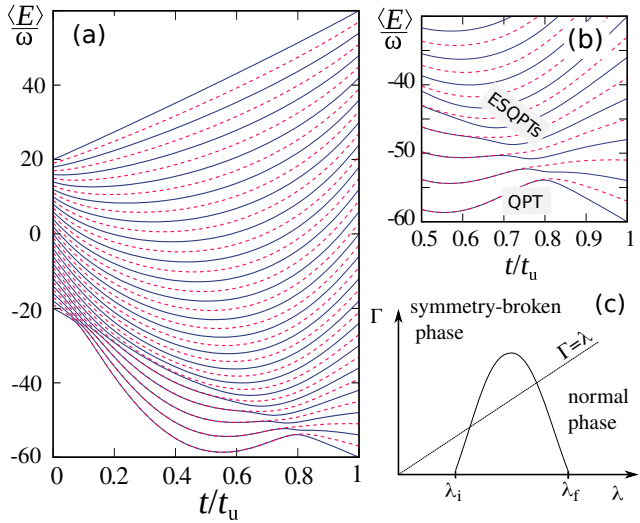}
	\caption{ Panel (a): Energy spectrum of the LMG model during the unitary stroke as a function of time. The blue full lines correspond to even parity while the red dashed lines to odd parity. Parameters of the model are $j=20, \ \lambda_{\rm i}=1, \ \lambda_{\rm f}=3, \ \bar{\Gamma}=15$. For $t=0$ and $t=t_u$ the energy spectrum is equidistant. Panel (b): Detail of the spectrum from panel (a), the QPT and a chain of ESQPTs are marked. Panel (c): A sketch of the dependence $\Gamma(\lambda)$. If the protocol is critical, then for certain values of $\lambda$ the system enters the symmetry-broken phase.
	}
	\label{fig7}
\end{figure}

The LMG model exhibits a quantum phase transition (QPT)  between the normal and the symmetry-broken phase at $\lambda(t)=\Gamma(t)$.
This ground-state QPT is accompanied by a chain of ESQPTs in the symmetry-broken phase, i.e., for $\Gamma(t)>\lambda(t)$~\cite{Gil78,Rib07,Cej15}.
As for $t=0$ and $t=t_u$ the system is in the normal phase, whenever the previous inequality of parameters is satisfied during the stroke, the critical point has been crossed.
In Fig.~\ref{fig7}(a) we present an example of how the energy levels evolve in the protocol given by Eqs.~\eqref{Eq:prot1} and~\eqref{Eq:prot2}.
The abrupt change of the ground state with $t$ corresponds to a QPT which is crossed twice.
In panel (b) a detail of the spectrum is shown indicating the QPT critical point and a chain of ESQPTs manifested by avoided crossings among the excited levels.
Because we work with relatively low temperatures, mostly the lowest lying states are populated and so only those ESQPTs in a close vicinity to the QPT critical point are relevant when driving the system through the critical region.
Panel (c) shows a sketch of the mutual dependence of $\Gamma$ and $\lambda$. 
The critical protocol corresponds to the situation when the system enters and leaves the symmetry-broken phase during the stroke.

A QPT as well as the associated ESQPTs are characterized by vanishing energy gaps between the neighboring energy levels in the $N\to\infty$ limit, which obviously represents an obstacle for quantum adiabatic driving.
In the following part we investigate how the finite-time quantum non-adiabatic driving through the critical region affects the heat engine performance.

\begin{figure}[t!]
	\centering
  \includegraphics[width=1\linewidth, angle=0]{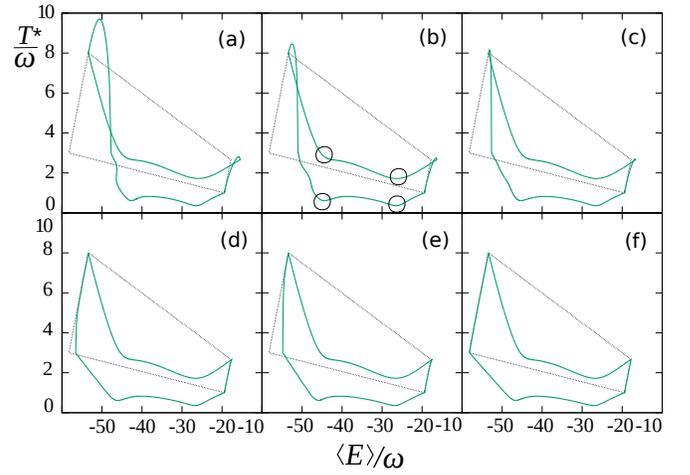}
	\caption{Quantum non-adiabatic evolution during the unitary strokes of the LMG model. The thermalization strokes are considered as perfect. The black dotted lines correspond to the fully thermalized cycle with no interaction as in Fig.~\ref{fig3}. The panels differ according to the finite value of the  time $t_u$ of the unitary strokes. Panel (a) $t_u=6 \omega^{-1}$, (b) $t_u=8 \omega^{-1}$, (c) $t_u=10 \omega^{-1}$, (d) $t_u=15 \omega^{-1}$, (e) $t_u=20 \omega^{-1}$ and (f) $t_u=100 \omega^{-1}$.
	The parameter value $\bar{\Gamma}=3$ guarantees that during the unitary stroke the system crosses the QPT.
	Other parameters are $j=20, \ \lambda_{\rm i}=1, \ \lambda_{\rm f}=3, \ T_c=1 \omega, \ \ T_h=8\omega, \ \gamma=0.1 \omega$.
	}
	\label{fig8}
\end{figure}

In Fig.~\ref{fig8} the cycles for several values of the duration of the unitary strokes $t_u$ are presented (we suppose the full thermalization in the corresponding strokes).
Parameters $\lambda_{\rm i}, \ \lambda_{\rm f}$ and $\bar{\Gamma}$ are selected in the way that the QPT is crossed during the unitary evolution.
In this case the reference temperature no longer approximates the thermodynamic one, however still some valuable information can be gained from its behavior during the cycle.
First, because of its definition~\eqref{Eq:VirtTempDef} the reference temperature inherently contains information on the structure of energy levels of the system.
Indeed, in all panels of Fig.~\ref{fig8} one can identify specific \lq bumps\rq\ in the unitary parts related to the the system entering or leaving the symmetry-broken phase [in panel (b) their position is pointed out explicitly by circles]. 
For relatively moderate $N/2=j=20$ these precursors may seem a little weak, nevertheless, it can be numerically proven that with growing $N$ these structures become much sharper.

\begin{figure}[t!]
	\centering
  \includegraphics[width=0.8\linewidth, angle=0]{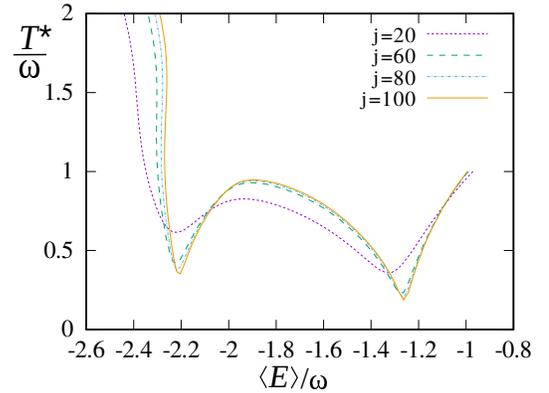}
	\caption{Quantum non-adiabatic evolution during the unitary stroke $1\to 2$ of the LMG model for different values of $j$. Duration of the stroke is $t_u=8 \omega^{-1}$ as in  Fig.~\ref{fig8}, panel (b). 
	Other parameters are $\bar{\Gamma}=3, \ \lambda_{\rm i}=1, \ \lambda_{\rm f}=3, \ T_c=1 \omega, \ \ T_h=8\omega, \ \gamma=0.1 \omega$.
	The dip indicating the QPT becomes sharper with growing $j$.
	}
	\label{fig9}
\end{figure}

The reason why the reference temperature forms a dip in $N\to\infty$ can be viewed from the following.
With methods used in Refs.~\cite{Ema04,Kop19} the Hamiltonian~\eqref{Eq:LMG} can be recast into a bosonic form.
After applying the Bogoliubov transform, the Hamiltonian is further mapped to a single harmonic oscillator mode  where the energy gap closes at the critical point.
All these transformations are unitary, hence conserve entropy.
So in the definition of the reference temperature~\eqref{Eq:VirtTempDef} we can replace the original Hamiltonian with the transformed one.
Therefore, keeping the entropy constant during the unitary evolution requires at closing energy gap a diverging $\beta^*$, i.e., a vanishing $T^*$, see Fig.~\ref{fig9}.

\subsection{Criticality and the work output}
\label{Sec:Work}
Now let us focus on the work output of the machine.
Surprisingly, already Fig.~\ref{fig8} can give us a hint on the performance of the engine.
For instance on panel (a) with a rather short time $t_u=6\omega^{-1}$ we can see an overshoot of the reference temperature above the value of the temperature of the hot reservoir ($T_h=8 \omega$).
In other words in this thermal stroke the reference temperature is not monotonously approaching the value $T_h$.
Based on Eq.~\eqref{Eq:heatflow} and the respective discussion below in Sec.~\ref{Subsec:RefTemp} it means that the heat current flows \textit{from} the WF \textit{to} the reservoir.
Thus, the hot bath is being heated up which obviously contradicts the functionality of the machine as a heat engine.
Similar overshoots (with the same consequences) can be noticed in panels (b) and (c) as well.

\begin{figure}[t!]
	\centering
  \includegraphics[width=0.8\linewidth, angle=0]{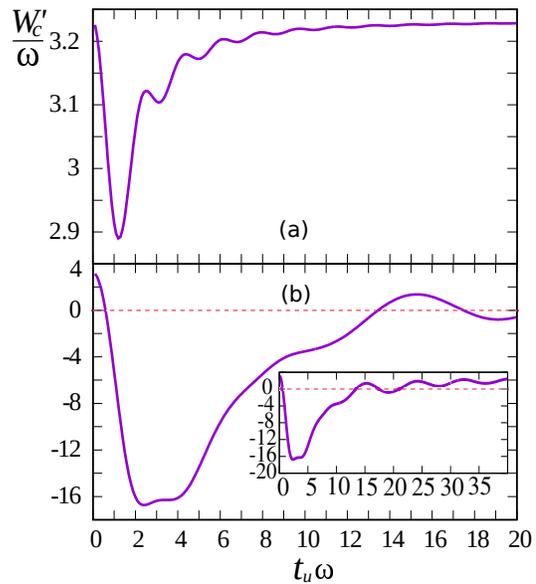}
	\caption{Extracted work in a cycle $W'_c$ as a function of the duration of the unitary stroke $t_u$.
	Panel (a): $\bar{\Gamma}=0.75$ (the QPT is not crossed).
	Panel (b): $\bar{\Gamma}=3$, same as in Fig.~\ref{fig7}. The QPT is crossed. The red dashed line marks zero work output level.
	The inset shows the result for longer time scale $t_u\in [0,40] \omega^{-1}$.
	Other parameters are $j=20, \ \lambda_{\rm i}=1, \ \lambda_{\rm f}=3, \ T_c=1 \omega, \ \ T_h=8\omega, \ \gamma=0.1 \omega$.
	}
	\label{fig10}
\end{figure}

The fact that in the cycle the heat is transfered from the WF to the hot reservoir is a result of an extremely inefficient unitary evolution in the preceding stroke.
Indeed, a large fraction of work was invested into population transfers so at the end the mean energy is greater than the thermal mean at $T_h$.
Figure~\ref{fig10} depicts the extracted work per cycle $W'_c$ as a function of $t_u$.
We stress again that the machine works as a heat engine only if $W'_c>0$.
In panel (a) the system remains in the normal phase during the unitary strokes, in panel (b) the critical point is crossed.

Both dependences have some common features.
For $t_u\to 0$ the situation corresponds to an abrupt quantum quench when the evolution is infinitely fast.
Then the final state is given simply by the distribution of the initial state in the final eigenbasis.
As in our protocol the initial and final eigenbases are the same [the  Hamiltonians for $t=0$ and $t=t_u$ have the same simple non-interacting form as in Eq.~\eqref{Eq:Ham}], after such a fast quench the populations are actually conserved.
Therefore, if initially in the thermal state, after the quench to $\lambda_{\rm f}$ the WF remains in the thermal state (with a different reference temperature) similarly as in Sec.~\ref{SubSecCollective}.
Therefore for extremely short times the work extracted reaches its maximum.
For growing $t_u$ the gain of work decreases very quickly, nevertheless for $t_u \gg 1$ (where the quantum adiabatic condition becomes more appropriately fulfilled) we retrieve the maximal work output.
However, panels (a) and (b) in Fig.~\ref{fig10} show a substantial difference.
In panel (a) where the QPT is \textit{not} crossed, the decrease in the work output is relatively shallow and stays in positive values.
Whereas in panel (b) depicting the situation where the critical point is crossed, $W'_c$ falls very deep into negative values which means that for large interval of $t_u$ the machine cannot work as a heat engine at all.
As can be anticipated from the quantum adiabatic theorem, the QPT and the associated ESQPTs (or better say even their precursor for finite $N$) form obstacles for quantum adiabatic driving which can easily bring the machine out of the useful operational mode.

Breaking of quantum adiabaticity and the resulting decrease in the amount of harvested work can also serve as an indicator of the QPT.
Since the dominant contribution to population transfers comes from the low-energy states (which are always the most populated ones), we estimate the proper criterion for adiabatic evolution from their behavior.
At the quantum-critical point the energy gap $\Delta E_{01}$ above the ground state closes.
For finite-size systems, the gap will therefore scale inversely with the system size and in particular for the LMG model this scaling is known to behave as $\Delta E_{01}\propto j^{-\frac{1}{3}}$~\cite{Dus04}.
However, also near the critical point one may already observe that the gap decreases with the system size.
In order to remain adiabatic as long as the critical point is not crossed, the driving time $t_u$ in Eqs.~\eqref{Eq:LMG},~\eqref{Eq:prot1} and \eqref{Eq:prot2} has to be carefully tuned.

Suppose we smoothly vary the parameter $\bar{\Gamma}$ in Eq.~\eqref{Eq:prot2}.
The critical value $\bar{\Gamma}_c$  corresponds to the setting where the critical point of the QPT has been exactly reached during the unitary strokes [the parabola in Fig.~\ref{fig7}(c) is tangent to the critical line $\Gamma=\lambda$].
Let us denote $\ket{\psi_0}$ the ground state and $\ket{\psi_1}$ the first excited state (from the same parity subspace)  of the Hamiltonian~\eqref{Eq:LMG}  at this critical point.
According to the quantum adiabatic theorem, it is necessary to evaluate the ratio $\scal{\psi_1}{\dot{H}|\psi_0}/(\Delta E_{01})^2$~\cite{Sch06} where $\Delta E_{01}$  is the energy gap between the two states $\ket{\psi_0}$ and $\ket{\psi_1}$ and $\dot{H}$ is the time derivative of the Hamiltonian~\eqref{Eq:LMG} which reads as
\begin{equation}
\dot{H}= -\frac{\lambda_{\rm f}-\lambda_{\rm i}}{t_u}\omega J_z-\frac{4\bar{\Gamma}}{N}\frac{1}{t_u}\Big(1-\frac{2t}{t_u} \Big)\omega J_x^2\,.
\label{Eq:TDerLMG}
\end{equation}
The~matrix element is evaluated for  $t=t_c$ and $\bar{\Gamma}=\bar{\Gamma}_c$ for which the quantum-critical point is reached.
The analytic expression for both $t_c$ and $\bar{\Gamma}_c$ can be obtained  from Eqs.~\eqref{Eq:prot1} and~\eqref{Eq:prot2} by setting $\Gamma(t_c)=\lambda(t_c)$  and requiring a single solution only.
After some algebraic manipulations one arrives at the expressions
 \begin{eqnarray}   
  \bar{\Gamma}_c&=&\frac{1}{4}(\sqrt{\lambda_{\rm i}}+\sqrt{\lambda_{\rm f}})^2, \label{eq:CritGam}
   \\
   t_c &=&\frac{t_u}{2} \sqrt{\frac{\lambda_{\rm i}}{\bar{\Gamma}_c}}. \label{Eq:tstar}
    \end{eqnarray}
Numerically we can prove the following scalings at the critical point
\begin{equation}
\scal{\psi_1}{J_z|\psi_0} \propto j^{\frac{1}{3}}, \qquad \scal{\psi_1}{J_x^2|\psi_0} \propto j^{\frac{4}{3}}.
\label{eq:ScalMatEls}
\end{equation}

From the previous paragraph it follows that $\scal{\psi_1}{\dot{H}|\psi_0}\propto j^\frac{1}{3}$. 
In total we obtain
\begin{equation}
 \frac{\scal{\psi_1}{\dot{H}|\psi_0}}{(\Delta E_{01})^2}\propto \frac{j}{t_u}\,,
\label{Eq:AdiabaticCond}
\end{equation}
so in order to make this term of order $1$ (hence break the adiabaticity at the critical point) we have to scale $t_u$~linearly with~$j$.
If $t_u$ is scaled with a larger power of $j$ (slow evolution) then the evolution remains adiabatic throughout.
On the other hand, if $t_u$ is scaled with a smaller power of $j$ (fast evolution) then the adiabatic condition breaks before the QPT.

While these arguments would be sufficient for a discussion of the low temperature regime, for finite temperatures a non-negligible fraction of the populations will be in the excited states already in the beginning of the unitary strokes.
Since in the thermodynamic limit $j\to\infty$ an effective harmonic oscillator description applies~\cite{Kop19}, the lower part of the spectrum will be equidistant as long as the critical point is not crossed.
Consistently, we numerically find that energy gaps and matrix elements between the first few excited states behave similarly.
Therefore, we expect the same arguments to hold also for the first few excited states, such that the above adiabaticity argument should generally hold.

\begin{figure}[t!]
	\centering
  \includegraphics[width=0.9\linewidth, angle=0]{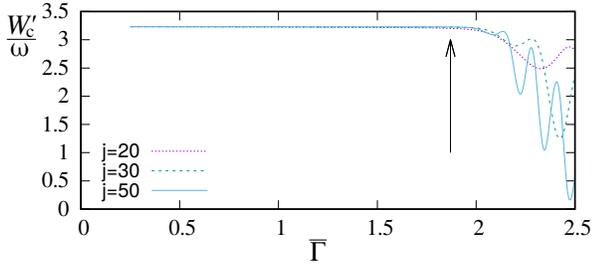}
	\caption{Extracted work in a cycle $W'_c$ as a function of parameter $\bar{\Gamma}$ from Eq.~\eqref{Eq:prot2}  for several values of $j$. The duration of the unitary stroke is $j$-dependent and set to $t_u= j\omega^{-1}$. 
	The arrow indicates the critical value $\bar{\Gamma}_c=1.87$ when the QPT is crossed.
	Other parameters are $\lambda_{\rm i}=1, \ \lambda_{\rm f}=3, \ T_c=1 \omega, \ \ T_h=8\omega, \ \gamma=0.1 \omega$.
	}
	\label{fig11}
\end{figure}

Indeed, we observe a breakdown of adiabaticity once the critical point is crossed in Fig.~\ref{fig11} where the work per cycle as a function of $\bar{\Gamma}$ is plotted for several values of $j$ while the time of unitary driving was always set to $t_u=j\omega^{-1}$.
Of course, the right-hand side of this dependence can be multiplied by any positive constant which would finely tune the protocol.
Even with this simple $j$ dependence one can see that crossing the critical value $\bar{\Gamma}_c$ (which is indicated by an arrow) leads to a decrease of the work output.
Putting it the other way round, simply by monitoring the performance of the engine one can distinguish whether the system has been driven across the QPT or not.

\subsection{Consequences and relevance to other known results}
\label{Subsec:Consequences}
Let us compare our results with the case where the presence of a QPT in the LMG model is reported to improve the efficiency~\cite{Ma17}.
First, the authors restrict themselves to a $J_z$-conserving version of LMG~\cite{Lip65} whose Hamiltonian remains diagonal in the eigenbasis of $J_z$ for any parameters $\lambda, \ \Gamma$ [similarly to our discussion of the non-interacting collective model~\eqref{Eq:Ham}].
Due to this fact, the criticality is associated with a real level crossing even for strictly finite $N$.
Indeed, the authors demonstrate the effect of the QPT already for $N=2$ which in the \lq standard\rq\ case would be highly problematic as the non-analytic features of the transition would appear in large-$N$ limit. 
Second, one should note that the cycle used by authors in Ref.~\cite{Ma17} is not a standard Otto cycle.
The difference is that in their case the evolution with $\lambda$ is supposed to be undergone in thermal equilibrium with the heat bath. 

A nice idea how to achieve the Carnot efficiency is to keep temperatures of both baths small $T_c, T_h\ll 1 \omega$ so the system is predominantly in the ground state, and set $\lambda_{\rm i}$ to the corresponding value of the real level crossing.
At this point, due to the thermal dissipators the ground state is doubly degenerate and therefore has non-zero entropy $S=\ln{2}$ which is used to extract maximal work~\cite{Ma17,Mas14}.
We should also stress that in this low-temperature setting the operational protocol can be fully replaced with a standard Otto cycle.
Indeed, as in the strokes where the internal parameters of the LMG Hamiltonians are being changed, the system (dominantly) stays in its ground-state.
Such a quantum adiabatic evolution is simply achieved by changing the parameters when detached from the heat bath due to the diagonal Hamiltonian as discussed in the previous paragraph.
Therefore comparison to our results is very relevant.

Authors in Ref.~\cite{Cha18} also model the WF with a $N=2$ LMG Hamiltonian of essentially a non-interacting type.
They demonstrate a similar feature as in Ref.~\cite{Ma17} and show that crossing the critical point creates an obstacle for the machine to extract work.
The reason for that can be intuitively seen as in the critical point the levels cross in this model (without mutual interaction), then after the stroke we obtain the state with swapped populations.
Similar to our case when the machine did not work as a heat engine, the final state after the unitary stroke had a higher energy then the respective thermal state of the heat bath.

Our results obtained with a $J_z$-violating LMG model with a standard QPT complements those achieved in Refs.~\cite{Ma17,Cha18}.
We can see that crossing (or even approaching) the critical point (for some large but still finite $N$) decreases the  amount of the extracted work $W'_c$.
Indeed, it can even go to negative values $W'_c<0$, however as the gap at the critical point is not really closed for finite $N$ we can still operate the machine slowly enough to make it work as a heat engine again.
The case when the system would undergo a thermal stroke when tuned to the critical point (in resemblance to Ref.~\cite{Ma17}) was not investigated within our driving protocol.
One can however guess that the positives (like long correlation length in the critical point which is known to increase the power output~\cite{Cam16}) would be traded off for the non-adiabatic losses in driving of the system as discussed in the Sec.~\ref{Sec:Work}.
However, it should be stressed that the derivation of a correct dissipator for near-degenerate systems is an open problem.

Our results are also relevant in context of Ref.~\cite{Cak16} where optimal working modes for and LMG with various internal parameters are studied.
The authors restrict themselves only on the system with small $N$ so their results are essentially not affected by any critical behavior.
However, if one considers large $N$ (which is beneficial in terms of power output as shown earlier), the negative effects of a QPT must be taken into account.



\section{Summary and conclusion}
In the first part we modeled the WF with a simple non-interacting spin system.
In the case of a single qubit we analytically reconstructed the cycle in the plane \lq mean energy vs. reference temperature\rq .
In a similar manner we provided a description of how the system approaches the limit cycle if finite-time thermal strokes are considered.

We demonstrated that for large $j$ and coherent dissipation the evolution in the \lq mean energy vs. reference temperature\rq\ plane is very similar to a single qubit.
The collectivity in dissipation significantly speeds up the thermalization process and so can boost the power output compared to the incoherent case.
We showed that the mean-field equation describing the superradiant burst (cf. Ref.~\cite{Gro82}) can be employed for the coherent dissipation provided that $j$ is sufficiently large.
The quadratic scaling $j^2$ of the power output was directly observed if the region $\ave{J_z}\approx 0$ was populated.
This required at least one of the baths to have relatively high temperature.
If one considers a model Hamiltonian of the type $H=\lambda J_z^2$ then this superradiant scaling should be easily observed even for small temperatures as the ground state already has the large Clebsch-Gordon coefficients. 

In the second part, we studied the Otto cycle with an interacting WF described by the Lipkin-Meshkov-Glick model.
Namely the effects of finite-time unitary strokes were investigated.
The effect of QPT and ESQPT precursors on the performance of the heat engine was generally negative.
If the unitary strokes were performed across the critical point then their overall duration had to be significantly slowed down in order to be able to extract work in the cycle.
This was the direct effect of increasing number of population changes in the parts of the spectrum where the levels get close to each other.
We showed that different regimes of unitary strokes (with and without a QPT) can be distinguished according to the work output per cycle.
A detailed discussion with relevance to other works can be found in Sec.~\ref{Subsec:Consequences}.

Taking into account the results presented in this paper, we can make the following statement.
During the equilibration strokes the collective coupling to the reservoirs is beneficial but during the unitary strokes the collective spin-spin interaction is detrimental to the power output.
Taking advantage of the collective effects therefore suggests to avoid critical points.

\section*{Acknowledgement}
The authors acknowledge the discussions with Wassilij Kopylov, Pavel Str\'{a}nsk\'{y}, Alexandre Roulet and Patrick Potts.
M.K. and G.S. gratefully acknowledge funding by the DFG (Projects
BR 1528/9-1 and BR 1528/8-2).
M.K. and P.C. acknowledge funding of the Charles University under Project No. UNCE/SCI/013.
M.K. was financially supported by the Swiss National Science Foundation (SNSF) and the NCCR Quantum Science and Technology.


\begin{thebibliography}{99}
 
 \bibitem{Ali79} R. Alicki, J. Phys. A: Math. Gen. \textbf{12}, L103 (1979).
 \bibitem{Kos13} R. Kosloff, Entropy \textbf{15}, 2100 (2013).
 \bibitem{Kos17} R. Kosloff, Y. Rezek,  Entropy \textbf{19}, 136 (2017).
\bibitem{Kos18} \textit{Thermodynamics in the Quantum Regime}, edited by F. Binder, L.A. Correa, C. Gogolin, J. Anders, G. Adesso (Springer International, 2019).
  
 \bibitem{Cak16} S. \c{C}akmak, F. Altintas,  \"{O}.E. M\"{u}stecaplıo\u{g}lu, Eur. Phys. J. Plus  \textbf{131}, 197 (2016).
 
 \bibitem{Scu03} M.O. Scully, M.S. Zubairy, G.S. Agarwal, H. Walther, Science \textbf{299}, 862 (2003).
 \bibitem{Ros14} J. Ro{\ss}nagel, O. Abah, F. Schmidt-Kaler, K. Singer, E. Lutz, Phys. Rev. Lett. \textbf{112}, 030602 (2014).
 \bibitem{Jar16} J. Jaramillo, M. Beau, A. del Campo, New J. Phys. \textbf{18}, 075019 (2016). 
 \bibitem{Nie18} W. Niedenzu, V. Mukherjee, A. Ghosh, A. G. Kofmann, G. Kurizki, Nature  Communications \textbf{9}, 165 (2018).
 
  
 \bibitem{Ros16} J. Ro{\ss}nagel, S.T. Dawkins, K.N. Tolazzi, O. Abah, E. Lutz, F. Schmidt-Kaler, K. Singer, Science \textbf{352}, 325 (2016).

 \bibitem{Mas19} G. Maslennikov, S. Ding, R. Hablutzel, J. Gan, A. Roulet, S. Nimmrichter, J. Dai, V. Scarani, D. Matsukevich, Nat. Commun. 10, 202 (2019).
  
 \bibitem{Kla19} J. Klatzow, J.N. Becker, P.M. Ledingham, C. Weinzetl, K.T. Kaczmarek, D.J. Saunders, J. Nunn, I.A. Walmsley, R. Uzdin, E. Poem, Phys. Rev. Lett. \textbf{122}, 110601 (2019). 
 
 \bibitem{Nis07} A.O. Niskanen, Y. Nakamura, J.P. Pekola, Phys. Rev. B \textbf{76}, 174523 (2007).
 
 \bibitem{Cam15} M. Campisi, J. Pekola, and R. Fazio, New J. Phys. \textbf{17}, 035012 (2015).
  
 \bibitem{Mar16} G. Marchegiani, P. Virtanen, F. Giazotto, M. Campisi, Phys. Rev. Applied \textbf{6}, 054014 (2016).
 
 
 \bibitem{Zha14} K. Zhang, F. Bariani,  P. Meystre, Phys. Rev. Lett. \textbf{112}, 150602 (2014).
 
 \bibitem{Gel15} D. Gelbwaser-Klimovsky, G. Kurizki, Sci. Rep. \textbf{5}, 7809 (2015).
 
  
  \bibitem{Gev92} E. Geva, R. Kosloff, J. Chem. Phys. \textbf{96}, 4 (1992).
 \bibitem{Fel04} T. Feldmann, R. Kosloff, Phys. Rev. E \textbf{70}, 046110 (2004). 
 \bibitem{Rez06} Y. Rezek, R. Kosloff, New J. Phys. \textbf{8}, 83 (2006).
 \bibitem{Sch08} T. Schmiedl, U. Seifert, Europhys. Lett. \textbf{81}, 20003 (2008).
 \bibitem{Esp10} M. Esposito, R. Kawai, K. Lindenberg, C. Van den Broeck, Phys. Rev. Lett. \textbf{105}, 150603 (2010).
 \bibitem{Esp10b} M. Esposito, R. Kawai, K. Lindenberg, C. Van den Broeck, Europhys. Lett. \textbf{89}, 20003 (2010).
 \bibitem{Fel12} T. Feldmann, R. Kosloff, Phys. Rev. E \textbf{85}, 051114 (2012).
 \bibitem{Bol12} F. Boldt, K.H. Hoffmann, P. Salamon, R. Kosloff, Europhys. Lett. \textbf{99}, 40002 (2012).
 \bibitem{Wan13} R. Wang, J. Wang, J. He, Y. Ma, Phys. Rev. E \textbf{87}, 042119 (2013).
 \bibitem{Wu14} F. Wu, J. He, Y. Ma, J. Wang, Phys. Rev. E \textbf{90}, 062134 (2014).
 \bibitem{Aba16} O. Abah, E. Lutz, Europhys. Lett. \textbf{113}, 60002 (2016).
 \bibitem{Ins16} A. Insinga, B. Andresen, P. Salamon, Phys. Rev. E \textbf{94}, 012119 (2016).
 \bibitem{Wie19} M. Wiedmann, J.T. Stockburger, J. Ankerhold, arXiv:1903.11368.

 
 
  \bibitem{Cam16} M. Campisi, R. Fazio, Nat. Commun. \textbf{7}, 11895 (2016).
  
  \bibitem{Vro18} H. Vroylandt, M. Esposito, G. Verley, Europhys. Lett. \textbf{120}, 3 (2018). 
 \bibitem{Nie18b} W. Niedenzu, G. Kurizki, New J. Phys. \textbf{20}, 113038 (2018). 
 \bibitem{Wat19} G. Watanabe, B.P. Venkatesh, P. Talkner, M.-J. Hwang, A. del Campo,  arXiv:1904.07811 
 
 \bibitem{Bea16} M. Beau, J. Jaramillo, A. del Campo,  Entropy \textbf{18}, 168 (2016).
  \bibitem{Cak19} B. \c{C}akmak, \"{O}.E. M\"{u}stecaplio\u{g}lu, Phys. Rev. E \textbf{99}, 032108 (2019).
 
 
 
 \bibitem{Sac11} S. Sachdev, \textit{Quantum Phase Transitions} (Cambridge Univ. Press, Cambridge, 2011).
\bibitem{Car11} \textit{Understanding Quantum Phase Transitions}, edited by L.D. Carr (CRC, Boca Raton, 2011).
\bibitem{Cej06} P. Cejnar, M. Macek, S. Heinze, J. Jolie, J. Dobe\v{s}, J. Phys. A: Math. Gen. \textbf{39}, L515 (2006). %
 \bibitem{Cap08} M. Caprio, P. Cejnar, and F. Iachello, Ann. Phys. (N.Y.)
323, 1106 (2008).
 \bibitem{Str16} P. Str\'{a}nsk\'{y}, P. Cejnar, Phys. Lett. A \textbf{380}, 2637 (2016).
 
\bibitem{Ma17} Y.-H. Ma, S.-H. Su, C.-P. Sun, Phys. Rev. E. \textbf{96}, 022143 (2017). 
\bibitem{Cha18} S. Chand, A. Biswas, Phys. Rev. E \textbf{98}, 052147 (2018).

  \bibitem{Kar11} O. Karlstr{\"o}m, H. Linke, G. Karlstr{\"o}m, and A. Wacker, Phys. Rev. B \textbf{84}, 113415 (2011).
  \bibitem{Vog11} M. Vogl, G. Schaller, T. Brandes, Ann. Phys. \textbf{326}, 2827 (2011).
  \bibitem{Sch16} G. Schaller, G.G. Giusteri, G.L. Celardo, Phys. Rev. E \textbf{94} 032135 (2016).

 \bibitem{Bin15} F.C. Binder, S. Vinjanampathy, K. Modi, J. Goold, New J. Phys. \textbf{17}, 075019 (2015). 
 \bibitem{Cam17} F. Campaioli, F.A. Pollock, F.C. Binder, L. C\'{e}leri, J. Goold, S. Vinjanampathy, K. Modi, Phys. Rev. Lett. \textbf{118}, 150601 (2017). 
 \bibitem{Le18}  T.P. Le, J. Levinsen, K. Modi, M.M. Parish, F.A. Pollock, Phys. Rev. A \textbf{97}, 022106 (2018). 
 \bibitem{Fer18} D. Ferraro, M. Campisi, G.M. Andolina, V. Pellegrini, M. Polini, Phys. Rev. Lett. \textbf{120}, 117702 (2018).
 \bibitem{And19} G.M. Andolina, M. Keck, A. Mari, M. Campisi, V. Giovannetti, M. Polini,
Phys. Rev. Lett. \textbf{122}, 047702 (2019).


 \bibitem{Dic54} R.H. Dicke , Phys. Rev. \textbf{93}, 99 (1954).
 \bibitem{Gro82} M. Gross, S. Haroche, Phys. Rep. \textbf{93}, 301 (1982).



 \bibitem{Lip65} H.J. Lipkin, N. Meshkov, N. Glick, Nucl. Phys. A \textbf{62}, 188 (1965).  
 \bibitem{Gil78} R. Gilmore, D.H. Feng, Nucl. Phys. A \textbf{301}, 189 (1978).
 \bibitem{Rib07} P. Ribeiro, J. Vidal, R. Mosseri, Phys. Rev. Lett. \textbf{99}, 050402 (2007).
 \bibitem{Cej15} P. Cejnar,  P. Str\'{a}nsk\'{y}, M. Kloc, Phys. Scr. \textbf{90}, 114015 (2015).
  
 \bibitem{Ali13} R. Alicki, M. Fannes, Phys. Rev. E \textbf{87}, 042123 (2013).
 \bibitem{Gel19} D. Gelbwaser-Klimovsky, W. Kopylov, G. Schaller, Phys. Rev. A \textbf{99}, 022129 (2019).
 
 \bibitem{Dav74} E.B. Davies, Comm. Math. Phys. \textbf{39}, 91 (1974).
 \bibitem{Spo78} H. Spohn, J. Math. Phys. \textbf{19}, 1227 (1978).

 
  \bibitem{Kie04} T.D. Kieu, Phys. Rev. Lett. \textbf{93}, 140403 (2004).
  \bibitem{Kie06} T.D. Kieu, Eur. Phys. J. D, \textbf{39}, 115-128 (2006).
  
  
  
  \bibitem{Nie00} M.A. Nielsen, I.L. Chuang, \textit{Quantum computation and quantum information} (Cambridge, Cambridge University Press, 2000).
 
   \bibitem{Bre02} H.-P. Breuer, F. Petruccione, \textit{Theory of Open Quantum Systems} (Oxford University Press, 2002).
   \bibitem{Wei12} U. Weiss, \textit{Quantum Dissipative Systems}, 4th ed. (World Scientific, 2012).
   
 \bibitem{Aga74}  G.S. Agarwal, \textit{Quantum statistical theories of spontaneous emission and their relation to other approaches} (Springer Tracts in Modern Physics, vol. 70, Springer, Berlin, 1974).
 
 
 

 
   
  \bibitem{Nov54} I.I. Novikov, J. Nucl. Energy \textbf{7}, 125 (1954).
  \bibitem{Cha57} P. Chambadal, Les centrales nucl\'{e}aires, pp. 41-58, Armand Colin, Paris (1957).
 \bibitem{Cur75} F.L. Curzon, B. Ahlbor, Am. J. Phys. \textbf{43}, 22 (1975).
 
 \bibitem{Dor18} K.E. Dorfman, D. Xu, J. Cao, Phys. Rev. E \textbf{97}, 042120 (2018).
 \bibitem{Abi19} P. Abiuso, V. Giovannetti, Phys. Rev. A \textbf{99}, 052106 (2019).
 

 \bibitem{Ema04} C. Emary, T. Brandes, Phys. Rev. A \textbf{69}, 053804 (2004).
 \bibitem{Kop19} W. Kopylov, G. Schaller, arXiv:1906.04260 
 
 
  \bibitem{Dus04} S. Dusuel, J. Vidal, Phys. Rev. Lett. \textbf{93}, 237204 (2004).
  
  \bibitem{Sch06} G. Schaller, S. Mostame, R. Sch\"{u}tzhold, Phys. Rev. A \textbf{73}, 062307 (2006).
 
 
 \bibitem{Mas14} E. Mascarenhas, H. Bragan\c{c}a, R. Dorner, M. Fran\c{c}a Santos, V. Vedral, K. Modi, J. Goold, Phys. Rev. E \textbf{89}, 062103 (2014).
 
 
 
 
 

 
\end{thebibliography}
\end{document}